\documentclass[twocolumn,amssymb, nobibnotes,showpacs, superscriptaddress, aps, prd]{revtex4}
\usepackage{amsmath}
\usepackage{graphicx}
\begin{document}
\title{Many-body dynamics of a  Bose system
with attractive interactions on a ring}
\author{Weibin Li}
 \affiliation{Wuhan Institute of Physics and Mathematics, The Chinese Academy of
Sciences, Wuhan 430071, People¡¯s Republic of China}
\affiliation{Graduate school, Chinese Academy of Sciences, Beijing
10080, People¡¯s Republic of China}
 \affiliation{Department of Physics,
Huazhong University of Science and Technology, Wuhan 430074,
People's Republic of China}
\author{Xiaotao Xie}
\affiliation{Department of Physics, Huazhong University of Science
and Technology, Wuhan 430074, People's Republic of China}
\author{Zhiming Zhan}
\affiliation{School of Physics and Information Engineering, Jianghan
University, Wuhan 430056, People's Republic of China}

\author{Xiaoxue Yang}
\affiliation{Department of Physics, Huazhong University of Science
and Technology, Wuhan 430074, People's Republic of China}
\begin{abstract}
We investigate the many-body dynamics of an effectively attractive
one-dimensional Bose system confined in a toroidal trap. The
mean-field theory predicts that a bright-soliton state will be
formed when increasing the interparticle interaction over a critical
point. The study of quantum many-body dynamics in this paper reveals
that there is a modulation instability in a finite Bose system
correspondingly. We show that Shannon entropy  becomes irregular
near and above the critical point due to quantum correlations. We
also study the dynamical behavior of the instability by exploring
the momentum distribution and the fringe visibility, which can be
verified experimentally by releasing the trap.
\end{abstract} 
\maketitle
\section{Introduction}
Bose-Einstein condensates (BECs) in quasi-one-dimensional systems
have attracted much more attention in recent years
\cite{gk,ss,ge,m,ccr,ccrr,gt,ag,sh,ks,ul,cc,kp,sr,ku,k,bb,bs},
because it, depending on the repulsive or attractive nature  of
two-body interaction, can form dark- or bright-soliton states,
respectively \cite{m,ccr,ccrr}. It is believed that atomic matter
bright solitons are of primary importance for developing concrete
applications of BECs in the future \cite{m}, so of particular
interests have been focused on the study of effectively attractive
BECs \cite{m,ccr,sh,ks,ku,k}. Using the Feshbach resonances
\cite{ik}, the coupling constant was tuned to a negative value and
both bright-soliton trains \cite{sh} and a single bright-soliton
\cite{ks} were realized experimentally.

The Gross-Pitaevskii mean-field  theory (MFT) is a powerful tool in
the study of BECs, which has successfully described an impressive
set of experiments \cite{kp,cc,sr}.  However it breaks down if
quantum many-body (QMB) correlations are important, such as in
quantum phase transitions \cite{gmehb,fwgf,jbcgz}. For
one-dimensional Bose system with attractive interaction, MFT
predicts the transition from a uniform state to a soliton state
\cite{ccr} at a critical point when increasing the absolute value of
the strength of the interaction. Recently, Kanamoto et.al. \cite{ku}
and Kavoulakis \cite{k} carefully checked the ground state and
low-lying excitation spectra near and above the transition. They
revealed that quantum correlations become crucial. The resulting
properties of QMB correlations, such as the singularity in the
transition is replaced by a crossover region, the condensate
fraction begins to decrease, etc. \cite{ku,k}, suggest that the MFT
should qualitatively be modified and QMB effects have to be
considered. On the other hand, most of the previous studies regard
QMB effects in the stationary situations, which attempt to beyond
MFT by using the diagonalization schemes \cite{ku,k}. Therefore it
is desirable to explore the quantum dynamics of this system. In the
present paper, we numerically study the quantum dynamics of N
interacting Bosons in a toroidal trap, which plays an important role
in the physics of trapped dilute gases \cite{bs}. As a function of
interaction strength, we calculate the Shannon entropy of the wave
packets to characterize the quantum dynamics. A modulation
instability at critical point is discovered, which is reflected by a
rapid increase of entropy. As a direct result, the interference
fringes, which occur after switching off the confinement and letting
the particles spread in a ballistic way, vanish near and above the
critical point.

The toroidal one-dimensional regime  can be realized when the
transverse dimensions are on the order of the healing length, and
the longitudinal dimension is much longer than the transverse ones
\cite{ccr,sh,ks,ku,k}. Experimentally, this geometry may be achieved
by an optical trap with Laguerre-Gaussian beams \cite{kss}. The
system is tightly confined in the radial direction so that the
energy-level spacings in this direction greatly exceed the
interaction energy. Thus N Bosons are confined on a  ring of radius
$R$ and cross section $S=\pi r^2$ with $r \ll R$. The Hamiltonian
for this system reads \cite{ccr,ku}
\begin{equation}
\label{h0} \hat{H}=\int_0^{2\pi}
d\theta[-\hat{\Psi}^\dag(\theta)\frac{\partial^2}{\partial\theta^2}\hat{\Psi}(\theta)+\frac{U}{2}\hat{\Psi}^\dag(\theta)\hat{\Psi}^\dag(\theta)\hat{\Psi}(\theta)\hat{\Psi}(\theta)]
\end{equation}
where $\hat{\Psi}(\theta)$ is the field operator, $U=8\pi aR/S$,
with $a$ being the s-wave scattering length, and $\theta$ denotes
the azimuthal angle. Here the length and the energy are measured in
units of $R$ and $\hbar^2 /2mR^2$. The normalization of
$\hat{\Psi}(\theta)$ is $\int|\hat{\Psi}(\theta)|^2d\theta=N$. The
Hamiltonian is characterized by a single dimensionless parameter
\cite{ccr,ku}
\begin{equation}
\label{para} \gamma \equiv \frac{UN}{2\pi}
\end{equation}
which is essential the ratio between the interaction energy and the
kinetic energy. The transition occurs when
$\gamma<\gamma_c\equiv-1/2$ \cite{m,ccr,ku,k}.

This paper is organized as follows. In sec.II, after numerically
obtaining the eigenenergies and eigenstates of the system, we study
the dynamical evolutions of Shannon entropy of the system. The
evolutions  show quasiperiodical  and irregular behaviors below and
above critical point correspondingly. The irregular one is
interpreted as much more number of virtually excited particles
involved in the dynamics due to strong interaction. In sec.III, we
discuss the momentum distribution of single-particle states. Above
critical point the system exhibits a statistical relaxation to a
steady distribution. The interference fringes of momentum, which can
be obtained experimentally by releasing the trap, are studied too.
We conclude in sec.IV

\section{Shannon entropy of the system}

In cylindrical coordinate expanding the field operator
$\hat{\Psi}(\theta,t)$ in terms of plane waves as
\begin{equation}
\label{sq}
\hat{\Psi}(\theta,t)=\frac{1}{\sqrt{2\pi}}\sum_{l}\hat{c}_l(t)e^{il\theta}
\end{equation}
with $l$ integer. Where $\hat{c_l}$ is annihilation operator of a
boson with angular momentum $l$. Then  Hamiltonian (\ref{h0}) can be
rewritten as
\begin{equation}
\label{eq1}
 \hat{H}=\sum_{l}l^2\hat{c_l}^\dag\hat{c_l}+\frac{\gamma}{2N}\sum_{kl;mn}\hat{c}_k^\dag\hat{c}_l^\dag\hat{c}_m\hat{c}_n
\end{equation}
 Expressing the single-particle
sates $|l\rangle$ according to their angular momentum
 $l=0,\pm 1,\pm 2, \dots$, a many-body basis state $|k\rangle$ can be represented as $|n_{ - L} ,
\cdots ,n_{ - 1} ,n_0 ,n_1 , \cdots ,n_L\rangle$ restricted to the
conservations of the total number of particles and the total angular
momentum
\begin{equation}
\label{mom}
 \sum_{l=-L}^{L} n_l=N, \sum_{l=-L}^{L} ln_l=0
\end{equation}
where $n_l$ denotes the number of bosons occupying the
single-particle sates $|l\rangle$, and $L$ is the cutoff of the
angular momentum. Since N bosons define the smallest spacing on a
ring, it is practical to make $L\approx N$ \cite{bs}. The  system
 is occupied by $m = 2L + 1$ single-particle
states.  With these assumptions, eigenvalues and eigenstates of
Hamiltonian (\ref{eq1}) can be solved directly by a smart method
\cite{wyx}(see Appendix).

\begin{figure}
 \includegraphics[width=9.5cm]{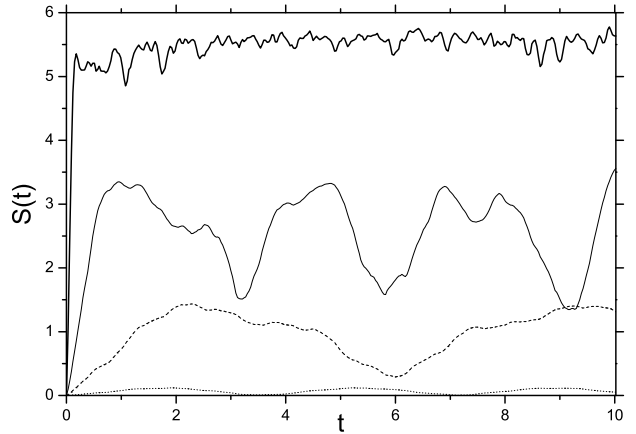}
 \caption{Entropy versus rescaled time t for $L=8,N=8$ and different $\gamma
 $. Dotted curve describes $\gamma=-0.14$. Dashed curve
corresponds to $\gamma=-0.54$. Thin solid curve represents
$\gamma=-1.62$ and thick solid curve represents $\gamma =-6.48$.}
\end{figure}

At first we introduce Shannon entropy to characterize the dynamical
behavior of the system, which has been used to manifest the
many-body dynamics of Tonks-Giraudeau gas \cite{bs}. The Shannon
entropy  is defined

\begin{equation}
\label{eq8} S(t) = - \sum_{k} |\Psi _k  (t)|^2\ln |\Psi _k (t)|^2
\end{equation}
Where $\Psi _k (t) = \langle k|\Psi (t) \rangle $ is the projection
of the wave function onto the many-body basis state $|k\rangle$. And
the time-dependent wave function reads
\begin{equation}
\label{eq7} |\Psi (t) \rangle = e^{ - iHt}|\Psi (0) \rangle
\end{equation}
Where $|\Psi (0) \rangle$ is state of the system at $t=0$.

Then, by varying $a$ via the Feshbach technology \cite{ik}, let us
consider an interesting case for which initially the system is
condensed on the ground state without interaction. After switch on
the interaction, the wave function spreads over all the basis.
\begin{figure}
 \includegraphics[width=9.5cm]{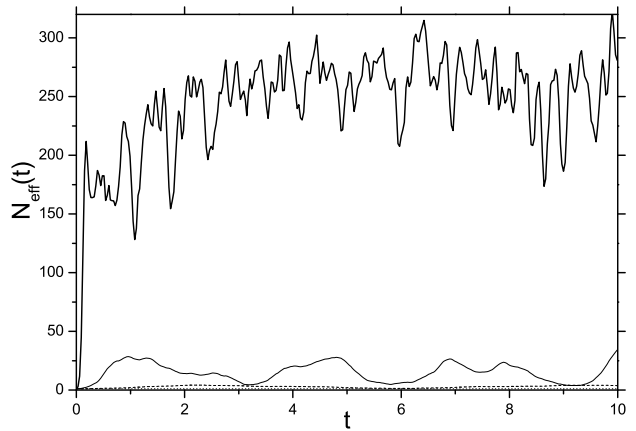}
 \caption{Time dependence of principal components for same parameters as fig.1.
 The  correspondences between line styles and  $\gamma$ are same as fig.1 too.}
\end{figure}

The time dependence of entropy is reported in Fig.1. For $0>\gamma
> \gamma_c$, the entropy oscillates quasiperiodically in time. For
$\gamma < \gamma_c$, the small oscillation is disappeared and the
entropy increases linearly to a saturation for a strong enough
interaction. For weak interactions  the system is condensed on the
ground state. The small deviations from ground state are suppressed
with time increasing \cite{ku,k}, which leads to the periodic
oscillations of the entropy. It also ensures the validity of MFT
with only a few modes \cite{ku,k}. When $\gamma$ is lowered
($|\gamma|$ is increased), the number of virtually excited particles
due to stronger interaction is increased \cite{ku,k}. This indicates
that much more basis states are involved in the dynamics. As a
result, the entropy is evolved with  a linear increase followed by a
saturation. It implies that the initial ground state decreases
exponentially while the basis excited in the dynamics increase
exponentially in time (see Fig.2). It can not be explained in the
MFT framework. To address the number of basis obviously, one may
calculate the principal components for wave packets in the basis
representation \cite{fi}. It is defined by the entropy
$N_{eff}(t)=\exp[S(t)]$, which is regarded as the effective number
of unperturbed many-body states in the dynamics.  A recent paper
about the two-body random interaction model \cite{fi} shows that in
a finite system, the number of basis states excited in the system is
finite too. Although Hilbert space is same for the system
independent of strength of interaction, the dynamics will depend
upon it. Stronger the strength of interaction, more basis involved
in the dynamics. So the finity of basis leads to the saturation of
the entropy. And the linear part of the time entropy may be
interpreted as the onset of the chaotic superpositions of the basis
states \cite{fi}.

\begin{figure}
 \includegraphics[width=9.5cm]{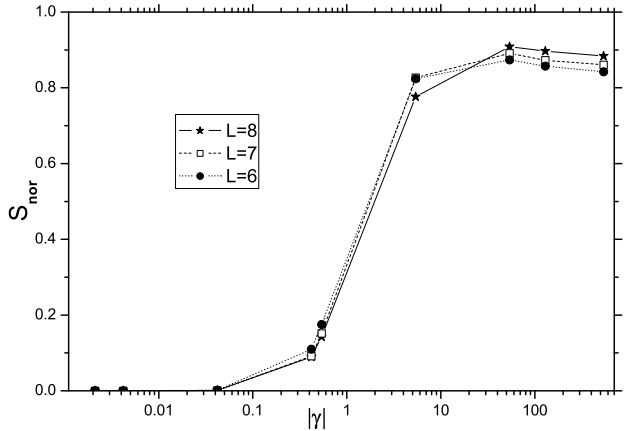}
 \caption{ Renormalized entropy for different $\gamma$ with
$N=8$ and $L=6,7,8$. The sharp increase  indicates the instability
region. Here we plot the absolute value of $\gamma$ for
convenience.}
\end{figure}

Although the complicacy of the quantum correlations near and above
the critical point \cite{ku}, the Shannon entropy is appropriate to
characterize the dynamical properties of the  system. To compare the
results for different $\gamma $, we calculate the rescaled entropy
$S_{nor} = S_{ave} / N_H $, where $S_{ave} $ is the mean value of
the entropy averaged over time after the saturation and $N_H $ is
the dimension of the Hilbert space, i.e. the total number of basis.
Fig 3 presents the results for $L=6,7,8$ and $N=8$. Here, different
$L$ can make us find the dependence of entropy on L. We can see that
the data are sensitive to the modulation instability. For $0>\gamma
> \gamma_c$, $\quad S_{nor} $ is stagnated to zero. This is same to the
MFT results, where bosons are almost condensed to the ground state.
Near the critical point, the virtual high-momentum single-particle
states are excited and the  system is influenced significantly by
the quantum fluctuations \cite{ku}. In this region, slight
variations of $\gamma $ alter the renormalized entropy obviously for
given $L$. For $\gamma \ll \gamma_c$, $S_{nor} $ reaches to a
plateau with altitude $\sim 0.85$. We guess that for the limitation
case $L\rightarrow \infty$, the renormalized entropy will reach to
unity corresponding to the situation that all of the basis states
have the same probabilities to be occupied.

\section{ Momentum distribution and fringe visibility}

Quantum dynamics characterized by the Shannon entropy is
investigated for different $\gamma$ in previous section. Similarly
to MFT, a modulation instability is founded at a critical point. But
near and above critical point, the irregular behaviors of quantum
dynamics can not be obtained by MFT due to quantum correlations.

QMB effects influence not only the entropy, but also the momentum
distribution, which can be observed in experiment. What we do in the
following is to address the influence.

Using the formal Fourier transforming, we obtain the momentum
function
\begin{equation}
\label{eq10} \hat {\chi }( \vec {p}) = \int dr\hat{\Psi}(\theta
)e^{ - i\vec {p}\cdot \vec {r}}
\end{equation}

Let us consider momentum along the y-axis $\vec {p} = p\cdot\hat {e}
_y $. We  get
\begin{equation}
\label{eq11} \hat {\chi }( \vec {p})= A\sum_{k}\hat{c}_k
\int\limits_0^{2\pi }d\theta e^{ik\theta }e^{ - ipR\sin \theta}
\end{equation}

Where A is a normalization constant. Note that the integration is
the Bessel function \cite{rhb} and substitute it in the equation.
Then
\begin{equation}
\label{eq12} \hat{\chi }(p) = A\sum_{n} \hat{c} _n  J_n (pR)
\end{equation}
Where $J_n (x)$ is the first kind Bessel function of order n. The
occupation number distribution is
\begin{equation}
\label{eq13} n(p,t) = \langle\Psi (t)| \hat {\chi } ^\dag (p)\hat
{\chi }(p)|\Psi (t) \rangle
\end{equation}
Using Eq.(\ref{eq7}) and Eq.(\ref{eq12}) , we obtain
\begin{equation}
\label{eq14} n(p,t) = \sum_{k,k'}\Psi _k^\dag\Psi _{k'}
\sum_{l,l'} {J_l }(pR)J_{l'} (pR) \langle k| \hat {c}_l^ \dag \hat
{c}_{l'}| k' \rangle
\end{equation}
With $ \langle k| \hat {c}_l^ \dag \hat {c}_{l'}| k' \rangle =
n_l^k \delta _{kk'} \delta _{ll'} $ where $n_l^k $ denotes the
number of particles with angular momentum $l$ in the basis
$|k\rangle$, we get
\begin{equation}
\label{eq15} n(p,t) = C\sum_{k} |\Psi _k | ^2\sum_{l} {J_l^2
(pR)n_l^k }
\end{equation}
Where C is a normalization constant.

\begin{figure}
 \includegraphics[width=9.5cm]{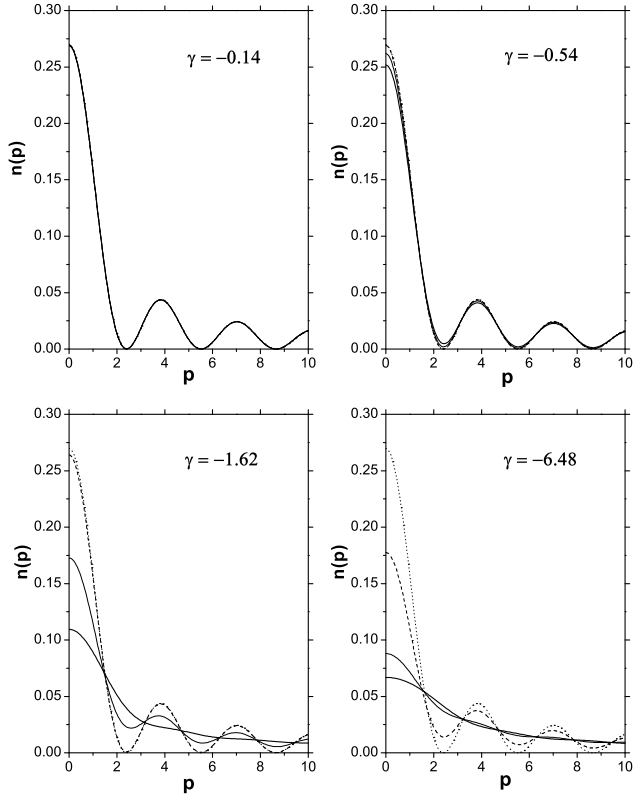}
 \caption{Particle density distributions in the momentum representation for four
parameters $\gamma$   at different times. Thicker curves describe
latter times. Dotted, Dashed, Thin solid, thick solid curves
correspond t=0, 0.1, 0.6, 1. }
\end{figure}

The momentum distributions for several $\gamma $ are shown in Fig.4.
 One can see that for weakly interacting
system the occupation number distribution will not be changed
obviously in time. But small perturbations lead to quasi-periodical
oscillation of the entropy (Fig.1). When the two-body interaction
strength exceeds the critical point, many high-momentum
single-particle states are excited. With time increasing, the
momentum distribution becomes flat and the condensate can not be
retrieval. A steady momentum distribution means the collapses of the
condensates. This suggests that we can use the fringe visibility to
address dynamical evolutions. The visibility is defined as \cite{bs}
\begin{equation}
\label{eq16} v(t) = \frac{|I_{\max } - I_{\min }|}{I_{\max } +
I_{\min } }
\end{equation}
Where $I_{\max } = n(p = 0,t),I_{\min } = n(p_0 ,t)$ and $p_0 $ is
the first zero of $J_0 $. The results are reported in Fig.5. The
periodical and irregular behaviors of the dynamics are showed for
some $\gamma$. For  $0>\gamma>\gamma_c$, quantum dynamics study
reveals that the visibility of a uniform ground state oscillates
with time (see inset of Fig.5), contrasting with the GP framework
where the visibility is unity \cite{ku,ccr}. For strong interaction,
the visibility is described by a fast decay followed by irregular
fluctuations. It can be used to distinguish the modulation
instability. So entropy, momentum distribution and visibility are
strictly consistent to characterize the dynamical behaviors of the
one-dimensional Bose system.

\begin{figure}
 \includegraphics[width=9.5cm]{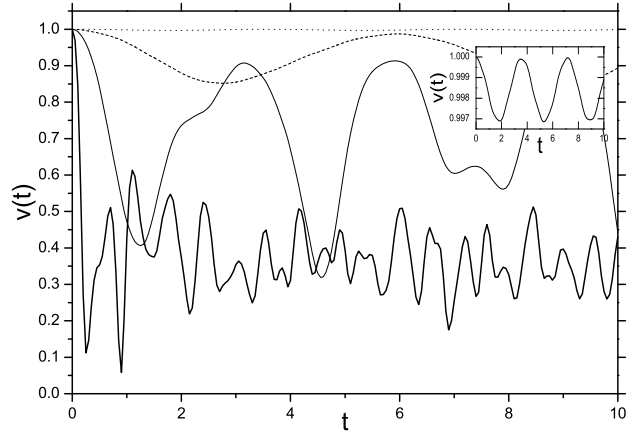}
 \caption{Fringe visibility of occupation number for different $\gamma$ .
 The curves correspond to same $\gamma$  as fig.1. Inset is the enlarging visibility for $\gamma=-0.14$ , which shows clearly periodic behavior.
}
\end{figure}

\section{Conclusion}

We studied the dynamical properties of attractive one-dimensional
Bose system with the QMB dynamics. Our main result is the onset of a
modulation stability at $\gamma_c$. We found that the quantum
dynamics of the system is distinguished by the periodical and
irregular behaviors for $\gamma$ below and above critical point
respectively.

Within the MFT, a phase transition between uniform state and bright
soliton is expected at a critical point \cite{ccr}. We investigate
the dynamical behaviors by varying the interacting strength. Far
bellow the critical point, the system is confined to the ground
state corresponding to a condensate. The entropy oscillates
quasiperiodically in this situation, which is connected with the
collapses and revivals of the condensate fraction (Fig.1 and Fig.2).
And the system is well described  by MFT. In the crossover region
the behaviors of entropy change gradually depending on $\gamma$. For
strong interaction regime, evolution of the entropy is characterized
by a fast linear increase followed by a saturation. This indicates
that the ground state is relaxed to steady states distribution and
the condensate is destroyed. The results of momentum distributions
and the fringe visibility confirm the modulation instability of the
system. They are of particular interest because they can be studied
experimentally.

The QMB effects, such as entropy, principal component, momentum
distribution and visibility have important meaning in the study of
the finite Bose system \cite{bs}.  The investigation in the context
demonstrates that QMB effects significantly influence the dynamical
behaviors and can help us to comprehend both the quantum evolution
of the system and the mechanism of the transition beyond MFT.
\begin{acknowledgments}
The work is supported in part by the NSF of China (Grant Nos.
60478029 and 10125419 ) and by the National Fundamental Research
Program of China, Grant No.2001CB309310. W.L. thanks Dr. Y. Wu for a
stimulating discussion.
\end{acknowledgments}
\appendix*
\section{energy spectrum and eigenstates of a Bose system on a ring}
In the following, we give an introduction about the algebra method
to calculate the eigenvalues of Hamiltonian (\ref{eq1}) based on
Ref. \cite{wyx}. One can see Ref.\cite{wyx} for a detail.

A many-body basis can be represented as
\begin{equation}
|n_{ - L} , \cdots ,n_{ - 1} ,n_0 ,n_1 , \cdots ,n_L\rangle
=\sqrt{\frac{1}{\prod_l
n_l!}}\prod_l(\hat{c}_l^{\dag})^{n_l}|vac\rangle
\end{equation}
where $|vac\rangle$ is the vacuum state satisfying
$\hat{c}_l|vac\rangle=0$. Hence, we denote the eigenstate of
Hamiltonian (\ref{eq1}) as
\begin{equation}
|\Psi\rangle=F(\hat{c}^\dag_{ - L} , \cdots ,\hat{c}^\dag_{ - 1}
,\hat{c}^\dag_0 ,\hat{c}^\dag_1 , \cdots ,\hat{c}^\dag_L)|vac\rangle
\end{equation}
where $F$ is the linear combination of the terms
$\prod_l(\hat{c}_l^{\dag})^{n_l}$. The eigenvalue equation
$H|\Psi\rangle=E|\Psi\rangle$ can be rewritten as
$EF|vac\rangle=HF|vac\rangle=[H,F]|vac\rangle$ due to
$H|vac\rangle=0$. Using $[\hat{c}_l,F]=\partial
F/\partial\hat{c}_l^\dag$, $[\hat{c}_l\hat{c}_m,F]=\partial^2
F/\partial\hat{c}_l^\dag\partial\hat{c}_m^\dag$, the polynomial
function $F$ satisfies the following differential equation
\begin{equation}
\sum_{l}l^2x_l\frac{\partial F}{\partial
x_l}+\frac{\gamma}{2N}\sum_{kl;mn}x_kx_l\frac{\partial^2F}{\partial
x_m\partial x_n}=EF
\end{equation}
where $x_l\equiv\hat{c}_l^{\dag}$. And the polynomial function
$F\equiv F(x_{ - L} , x_{ - 1} ,x_0 ,x_1 , \cdots ,x_L)$ is the
linear combinations of the terms $\prod_lx_l^{n_l}$. Now, the
eigenvalues problem is reduced to seek polynomial solutions to a
second order liner differential equation. Utilizing this method, it
is not a hard work to use a MATHEMATICA program to calculate all the
eigenvalues and eigenstates of Hamiltonian (\ref{eq1}) for finite
$N$ and $L$.


\begin{thebibliography}{}

\bibitem{gk}A. G$\ddot{o}$rlitz, J.M. Vogels, A.E. Leanhardt, C.
Raman, T.L. Gustavson, J. R. Abo-Shaeer, A.P. Chikkatur, S. Gupta,
S. Inouye, T. Rosenband, and W.Ketterle, Phys. Rev. Lett.
\textbf{87},130402(2001).

\bibitem{ss}
F. Schreck, L. Khaykovich, K. L. Corwin, G. Ferrari, T. Bourdel,
J. Cubizolles, and C. Salomon, Phys. Rev. Lett. \textbf{87},
080403 (2001)

\bibitem{ge} Markus Greiner, Immanuel Bloch, Olaf Mandel, Theodor W. H\"{a}nsch, and Tilman
Esslinger, Phys. Rev. Lett.\textbf{ 87}, 160405 (2001)

\bibitem{m}
P. Meystre,Atom Optics, AIP press, New York, 2001, and references
therein.

\bibitem{gmehb} M. Greiner, O. Mandel, T. Esslinger, T.W. H\"{a}nsch, and
I. Bloch, Nature (London) 415, 39 (2002).

\bibitem{fwgf}
M. P.A. Fisher, P. B. Weichman, G. Grinstein, and D. S. Fisher,
Phys. Rev. B 40, 546 (1989).

\bibitem{jbcgz}
 D. Jaksch, C. Bruder, J. I. Cirac, C.W.
Gardiner, and P. Zoller, Phys. Rev. Lett. 81, 3108 (1998).

\bibitem{ccr}
L.D. Carr, C.W. Clark, and W.P. Reinhardt, Phys. Rev. A
\textbf{62},063611(2000).

\bibitem{ccrr}L.D. Carr, C.W. Clark, and W.P. Reinhardt, Phys. Rev. A
\textbf{62},063610(2000).

\bibitem{gt}
M. D. Girardeau, E. M. Wright, and J. M. Triscari, Phys. Rev. A
\textbf{63}, 033601 (2001)

\bibitem{ag}
G. E. Astrakharchik and S. Giorgini, Phys. Rev. A \textbf{66},
053614 (2002);

\bibitem{sh}K.E. Strecker, G.B. Partridge, A.G. Truscott, and R.G. Hulet, Nature
,London \textbf{417}, 150(2002).

\bibitem{ul}M.Ueda and A.J. Leggett, Phys. Rev. Lett. \textbf{83},
1489 (1999)


\bibitem{ik}S. Inouye,M.R. Andrews, J. Stenger, H.-J. Miesner, D.M.
Stamper-Kurn, and W. Ketterle, Nature(London)
\textbf{392},151(1998).

\bibitem{ks}L. Khaykovich, F. Schreck, G.
Ferrari, T. Bourdel, J. Cubizolles,L.D. Carr, Y. Castin, and C.
Salomon, Science \textbf{296},1290(2002).


\bibitem{cc}L.D. Carr and Y. Castin, Phys. Rev. A \textbf{66}, 063602 (2002).


\bibitem{kp}U.Al Khawaja, H.T.C. Stoof, R.G. Hulet, K.E. Strecker, and
G.B.Partridge, Phys. Rev. Lett. \textbf{89}, 200404(2002)

\bibitem{sr}L. Salasnich, A.Parola, and L. Reatto, Phys. Rev. A \textbf{66},
043603(2002).

\bibitem{ku}R. Kanamoto, H. Saito, and M. Ueda, Phys. Rev. A \textbf{67},
013608(2003)

\bibitem{k}G.M. Kavoulakis, Phys. Rev. A \textbf{67}, 011601(\textbf{R})(2003)

\bibitem{bb}G.P.Berman, A. Smerzi, and A. R. Bishop, Phys. Rev.Lett. \textbf{88},
120402 (2002), and references therein.

\bibitem{bs}G.P.Berman,F.Borgonovi, F.M.Izrailev,A. Smerzi, Phys. Rev. Lett. \textbf{92}, 030404(2004)

\bibitem{kss} T.Kuga, Y. Torii, N. Shiokawa, T. Hirano, Y. Shimizu,
and H. Sasada, Phys. Rev. Lett. \textbf{78}, 4713(1997).


\bibitem{wyx}Y. Wu, X.Yang, and Y. Xiao, Phys. Rev. Lett. \textbf{86}, 2200(2001)

\bibitem{fi}V.V. Flambaum and F.M. Izrailev, Phys. Rev. E \textbf{64},036220
(2001).

\bibitem{rhb} K. F. Riley, M. P. Hobson and S. J. Bence, Mathematical methods
for physics and engineering: A comprehensive guide(second edition),
Cambridge University Press, Cambridge (UK),2002. See Page 574.
\end{thebibliography}
\end{document}